# Oxygen vacancy mediated cubic phase stabilization at room temperature in pure nano-crystalline Zirconia films: A combined experimental and first-principles based investigation

Parswajit Kalita[1,*], Shikha Saini[1], Parasmani Rajput[2], S. N. Jha[2], D. Bhattacharyya[2], Sunil Ojha[3], Devesh K. Avasthi[4], Saswata Bhattacharya[1] and Santanu Ghosh[1]

[1]*Department of Physics, Indian Institute of Technology Delhi, New Delhi -110016, India.*

[2]*Atomic & Molecular Physics Division, Bhabha Atomic Research Center, Mumbai – 400085, India.*

[3]*Inter University Accelerator Center, New Delhi – 110067, India.*

[4]*Amity Institute of Nanotechnology, Amity University, Noida – 201313, India.*

## Abstract

We report the formation of cubic phase, under ambient conditions, in thin films of Zirconia synthesized by electron beam evaporation technique. The stabilization of the cubic phase was achieved without the use of chemical stabilizers and/or concurrent ion beam bombardment. Films of two different thickness (660 nm, 140 nm) were deposited. The 660 nm and 140 nm films were found to be stoichiometric ($ZrO_2$) and off-stoichiometric ($ZrO_{1.7}$) respectively by Resonant Rutherford back-scattering spectroscopy. While the 660 nm as-deposited films were in the cubic phase, as indicated by X-ray diffraction and Raman spectroscopy measurements, the 140 nm as-deposited films were amorphous and the transformation to cubic phase was obtained after thermal annealing. Extended X-ray absorption fine structure measurements revealed the existence of Oxygen vacancies in the local structure surrounding Zirconium for all films. However, the amount of these Oxygen vacancies was found to be significantly higher for the amorphous films as compared to the films in the cubic phase (both 660 nm as-deposited and 140 nm annealed films).

The cubic phase stabilization is explained on the basis of suppression of the soft $X_2^-$ mode of vibration of the Oxygen sub-lattice due to the presence of the Oxygen vacancies. Our first-principles modeling under the framework of density functional theory shows that the cubic structure with Oxygen vacancies is indeed more stable at ambient conditions than its pristine (without vacancies) counterpart. The requirement of a critical amount of these vacancies for the stabilization of the cubic phase is also discussed.

*Correspondence email: phz148112@physics.iitd.ac.in

## 1. Introduction

The oxide ceramic zirconium oxide ($ZrO_2$), commonly referred to zirconia, exists in the monoclinic phase ($P2_1/c$ space-group) at room temperature. It reversibly gets transformed into the tetragonal phase ($P4_2/nmc$ space-group) and cubic phase (cubic fluorite, $Fm3m$ space group) at ~ 1170ºC and ~ 2370ºC respectively [1]. Zirconia, in its cubic and tetragonal form, has attracted immense attention as a key engineering material because of its excellent mechanical, thermal and optical properties. It possesses high hardness as comparable to that of steel and other alloys [2], exceptional radiation tolerance [3,4], low thermal conductivity, highest dielectric constant amongst metallic oxides [5], low light absorption and high refractive index [6]. In view of this, zirconia has been used for a wide variety of applications ranging from thermal barrier and wear resistance coatings [7,8], protective layer for optical mirrors and filters [6], high temperature solid oxide fuel cells [9,10] to oxygen sensors [9], catalyst support [11] etc. It is also a promising candidate for use as the inert material matrix in inert matrix fuels (IMFs) in nuclear reactors [4,12] and as a high-K gate dielectric in metal-oxide semiconductor devices [13]. However, since zirconia is not inherently stable in the cubic phase at room temperature, the application of zirconia crucially depends on the stabilization of the cubic phase at room temperature. This is generally achieved by the addition of lower valence stabilizer oxides, such as $Y_2O_3$, MgO etc., which creates oxygen vacancies (in the zirconia structure) to energetically favor the cubic phase [14-16]. Using this method, cubic zirconia is normally obtained by the addition of 8 to 20 mol% of these stabilizer oxides [14,15]. However, the properties of such 'stabilized' zirconia deteriorate because of the presence of such large quantities of stabilizing oxides (viz. chemical stabilizers) [17]. The latter is detrimental for practical applications [16]. It is therefore desirable to stabilize the cubic phase at room temperature without the use of any chemical stabilizers.

It is now well-known that the phase stability of materials under ambient conditions can be controlled via the 'nanosize' effect [16] – it's been illustrated for various materials that the decrease in the crystallite size leads to the formation of metastable crystal structures/phases that are only possible for their bulk counterparts under high temperature and pressure [18,19]. However, it remains a challenge to synthesize and maintain the high temperature and pressure polymorphs of materials, at ambient conditions, with crystallite sizes greater than the critical sizes as a result of the 'nanosize' effect alone.

In the recent past, ion beam assisted deposition (IBAD) has been used for synthesizing zirconia thin films that are stable in the cubic phase (without the presence of any chemical stabilizers) at room temperature, with an average crystallite size of ~ 8 – 10 nm [15,16,20]. The IBAD process combines evaporation with simultaneous ion beam bombardment and the crystallinity, vacancy formation, morphology, density etc. of the prepared films can be modulated by controlling the ion bombardment [15]. Wang et al. [16] reported cubic $ZrO_2$ films (at room temperature) synthesized by IBAD, using the ion beam of nitrogen and argon mixture at an energy of 500 eV, and the stabilization of cubic phase was attributed to the implanted nitrogen. On the other hand, Soo et al. [20] attributed oxygen vacancies to be responsible for stabilization of the cubic phase at room temperature in zirconia films prepared by IBAD, with the same ion species and energy. It is, therefore, worth mentioning two important points here from these previous studies: (i) the presence of implanted nitrogen in the films (in Ref. [20]) and (ii) the implanted nitrogen may also play a role (along with the oxygen vacancies) in the stabilization of the cubic phase. Therefore, in these IBAD prepared cubic zirconia films, it is important to note that even though the films were prepared without any chemical stabilizers, they were still not completely 'pure' due to the presence of the implanted nitrogen (viz. 'dopants'). The formation of the cubic phase at room temperature in stabilizer free zirconia films prepared by spin-coating (followed by thermal annealing) has also been reported [21].

In this article, we report, presumably for the first time, pure (i.e. without any chemical stabilizers and/or dopants) nano-crystalline zirconia thin films synthesized by standard electron-beam assisted thermal evaporation technique. The films were found to be in the cubic phase at room temperature and pressure (i.e. under ambient conditions) with an average crystallite size of ~ 10 nm. Note that the formation of cubic phase has been achieved without any 'external perturbations', i.e., (i) the use of chemical stabilizers, (ii) concurrent ion beam bombardment, (iii) thermal annealing or any other post deposition treatments, that can drive the system towards the cubic phase (whether it be stable or metastable). It's to mention here that the crystallite size in the films is greater than the reported critical size (~ 2 nm) [22] for cubic phase stabilization (via the 'nanosize' effect) in stabilizer free zirconia. Thus, the 'nanosize' effect alone does not seem to be a probable explanation for the cubic phase stabilization in the present case. The stabilization of the cubic phase at ambient conditions is primarily attributed to the hardening of the soft $X_2^-$ mode of vibration caused by the presence of oxygen vacancies. A critical amount of oxygen vacancies may however be required for the phase stabilization. From an application point of view, the formation of cubic phase in these films is quite encouraging due to the cost-effectiveness and simplicity of electron-beam evaporation as compared to IBAD. It is to be noted that the formation of cubic phase (at room temperature) has also been reported for pure nano-crystalline zirconia powders prepared by the inexpensive high energy ball-milling technique [23]. However firstly, it is difficult to produce adherent and uniform coating/layers/films (for various practical applications) using powders and secondly, the cubic phase nano-crystallites/domains of ~ 2 – 3 nm size in these powders [23] are significantly lower than the average crystallite size of the reported films in this work.

## 2. Experimental & theoretical details:

### 2.1 Fabrication of nano-crystalline zirconia films

Stabilizer free nano-crystalline zirconia thin films are synthesized by electron beam evaporation technique. For this purpose, 99.9% pure monoclinic zirconia powder (Specpure) is evaporated (after being compacted into pellet form) using an electron beam onto cleaned silicon <100> substrates at a rate of 0.2 – 0.6 Å/s. Films of two different thicknesses are fabricated by varying the time of deposition. The depositions are performed in an ultra-clean vacuum environment with a base pressure and working pressure of ~ $2 \times 10^{-8}$ Torr and ~ $5 \times 10^{-7}$ Torr respectively. No concurrent ion beam bombardment or external heating of the substrates is provided at any stage of the deposition process. Some of the films of lower thickness are subjected to subsequent thermal annealing in air for 30 minutes at a temperature of 500º Celsius in a quartz tube furnace. The as-deposited films of lower thickness, annealed films (of lower thickness) and as-deposited films of higher thickness are hereafter referred to as S1, S2 and S3 respectively.

## 2.2 Characterization

Resonant Rutherford backscattering spectrometry (RRBS) with a 3.05 MeV $He^{2+}$ beam is used to determine the thickness, composition and stoichiometry of the zirconia thin films. The $^{16}O(\alpha, \alpha)^{16}O$ reaction with a resonance energy of 3.042 MeV is chosen to measure the oxygen concentration in the samples since the scattering cross-section is significantly enhanced in case of resonance reactions [24]. A silicon surface barrier detector set at an angle of 165º from the incident beam direction is used to detect the backscattered alpha particles. The thickness and composition of the thin films is obtained from the simulation of the experimental RRBS data using RUMP [25-27] software. Energy dispersive X-ray spectroscopy (EDX) has also been performed to determine the elemental compositions of the films.

The phase identity and crystalline structure of the films is investigated by X-ray diffraction (XRD) and Raman spectroscopy. XRD measurements are carried out with $Cu_{K\alpha}$ radiation ($\lambda = 1.5406$ Å) in standard 2θ geometry, with the incidence angle fixed at 1º, using a Philips X'Pert Pro diffractometer. The average crystallite sizes are estimated from the XRD data, after correcting for instrumental broadening, using the Scherrer equation with a shape factor of 0.9. The Raman spectra are recorded using a Raman confocal microscope (Horiba LabRAM HR Evolution) with an Argon ion laser ($\lambda = 514$ nm) used as the excitation source.

Extended X-ray Absorption Fine Structure (EXAFS) is used to probe the local structural order/environment in the films. The EXAFS measurements are carried out at the zirconium (Zr) K-edge in conventional fluorescence mode (using a Vortex detector) at the Scanning EXAFS beamline (BL-9) of the Indus-2 Synchrotron Source at the Raja Ramanna Centre for Advanced Technology (RRCAT), India. The energy range is calibrated using zirconium metal foils. The EXAFS data is analyzed using FEFF 6.0 code [28], which includes background reduction and Fourier transform to derive the $\chi(R)$ versus R spectra from the absorption spectra (using ATHENA software), generation of the theoretical XAFS spectra starting from an assumed crystallographic structure and finally fitting of experimental data with the theoretical spectra using ARTEMIS software [29].

### 2.3 Theoretical methodology

Spin-polarized density functional theory (DFT) calculations are performed, by using Vienna *ab-initio* Simulation Package (VASP) [30] with projected augmented wave (PAW) [31] potential. The Perdew-Burke-Ernzerhof (PBE) exchange correlation functional [32] within generalized gradient approximation (GGA) is used in all the calculations. The cut-off energy of 600eV is chosen for the plane wave basis set. The tolerance energy for the convergence is set to 0.01 meV to achieve the self-consistency in total energy. All the electronic configurations are fully relaxed until the ionic forces were smaller than 0.001 eV/Å using conjugate gradient minimization. The supercell size is kept 25 Å to prevent the interaction between the adjacent periodic images of the nanocluster. Further, we have employed *ab initio* atomistic thermodynamics to identify thermodynamically the most favorable configuration/structure at ambient conditions (T=300K, $p_o$ = 1atm) [33-36]. The size of the nanocluster (used in our simulations) is taken considering the thickness of the films as in our experiment. In the simulations, to mimic the S2 film, we have modeled the spherical nanocluster of 46 atoms ($Zr_{14}O_{32}$), from bulk cubic zirconia, whereas the nanocluster of 131 atoms ($Zr_{43}O_{88}$) has been constructed to correspond the S3 film.

### 3. Results

The RRBS data of S1, S2 and S3 films are shown in Figure 1. The thickness and composition of the films obtained by fitting the experimental RRBS data are summarized in Table 1. It can be seen that the S3 film, being significantly thicker than S1 and S2 films, is also stoichiometric (O/Zr ratio = 2), whereas the S1 and S2 films are oxygen deficient and hence non-stoichiometric (O/Zr ratio = 1.7). On annealing the thinner films, the thickness, composition and stoichiometry remained unchanged. Moreover, RRBS signal corresponding to any element other than zirconium and oxygen is not observed. This is a validation to rule out the presence of any impurity in these films. EDX measurement also reveals the presence of only zirconium and oxygen in these films (not shown here).

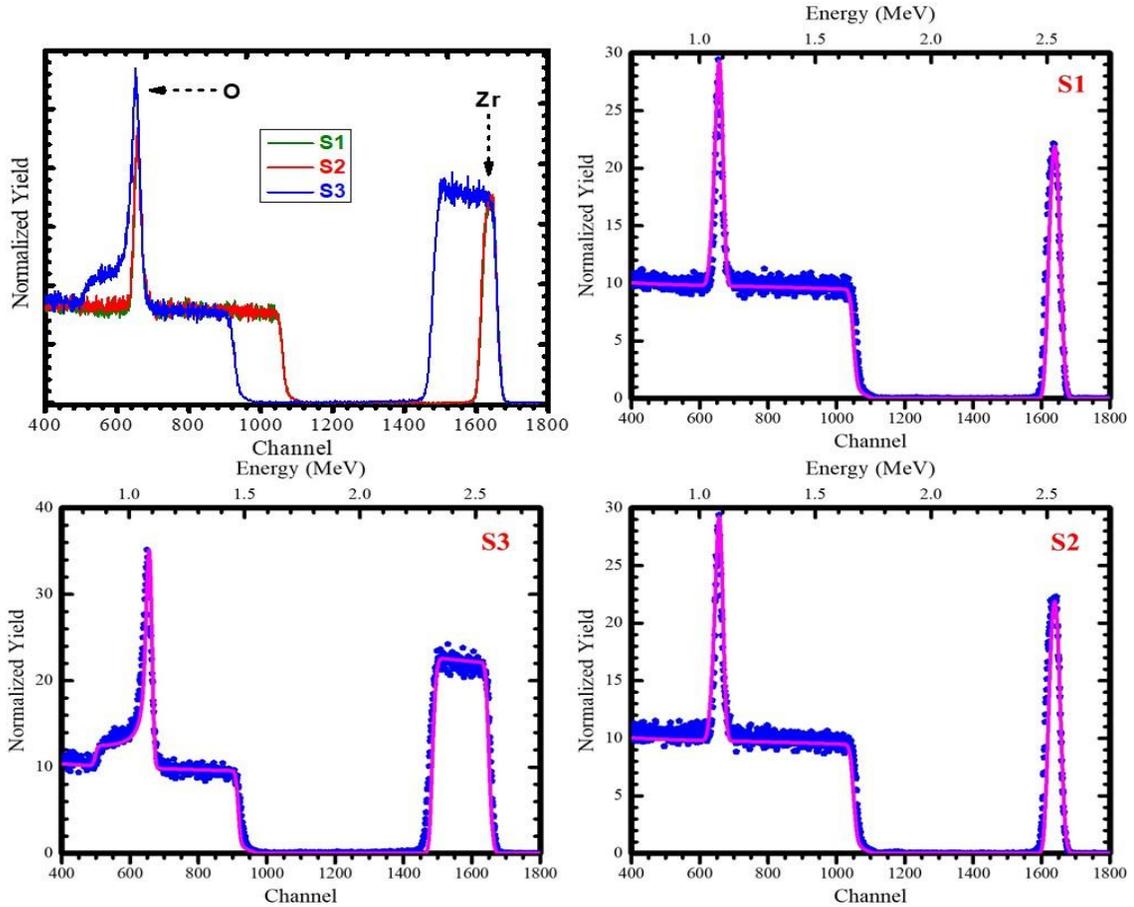

**Figure 1:** Experimental and fitted RRBS spectra of S1, S2 and S3 films indicating the presence of Zirconium (Zr) and Oxygen (O).

Figure 2 shows the XRD pattern of the zirconia films. For the S3 film, the presence of the monoclinic phase is immediately ruled out, as the most intense peaks corresponding to the monoclinic phase are not present. On comparing the XRD pattern of the S3 film with ICDD PDF cards, we find the powder diffraction patterns of cubic phase [PDF# 49-1642] of zirconia.

| Sample | Thickness | Composition |
|---|---|---|
| S1 (as-deposited) | 140 nm | $ZrO_{1.7}$ |
| S2 (annealed S1) | 140 nm | $ZrO_{1.7}$ |
| S3 (as-deposited) | 660 nm | $ZrO_2$ |

**Table 1:** Thickness and composition of S1, S2 and S3 films as determined by the fitting of experimental RRBS data.

However, it should be mentioned here, that for the two phases (viz. cubic and tetragonal) the axial ratio, *c/a*, is very similar (cubic [1.00] and tetragonal [1.01 - 1.02] [37,38]). It is, therefore, difficult (especially with the broadening of the peaks due to the nano-crystalline structures) to identify if the XRD pattern corresponds to a tetragonal or a cubic phase of zirconia. But, the region around the (400) diffraction maxima occurring at ~ $2\theta = 74º$ is a signature peak to distinguish these two phases. Note that in case of the tetragonal phase [37,38], the (400) peak of cubic zirconia splits into the two characteristic (004) and (400) peaks. The separation of the (004) and (400) peaks of the tetragonal phase is usually more than 1º [38] and thus provide evidence to identify if the phase is either cubic or tetragonal even in nano-crystalline zirconia. In our case, we find in the XRD spectrum of S3 (see Figure 2), only one (400) peak at (~ $2\theta = 74.5º$). This confirms the formation of cubic zirconia only. Moreover, the absence of the (112) peak at ~ $2\theta = 43º$ further validates the presence of only cubic phase, where this (112) peak is another characteristic peak for tetragonal phase. Therefore, all the characteristic features to establish the formation of only the cubic phase are present in this S3 film. It must be mentioned here that the present observation of the formation of only the cubic phase in films prepared solely by electron beam evaporation (i.e. without concurrent ion beam bombardment) is in sharp contrast to previous reports [15,16] where, the co-existence of different zirconia phases was observed.

Unlike S3, the S1 film is found to be amorphous (see Figure 2). Crystalline phase is obtained after thermal annealing (see S2 film, Figure 2). This phase is again verified to be the cubic phase of zirconia (i.e. absence of (112) peak and appearance of a single (400) peak in addition to the characteristic cubic (111), (200), (220), (311) peaks). The average crystallite size in S2 and S3 films are calculated to be ~ 13 nm and ~ 10 nm respectively.

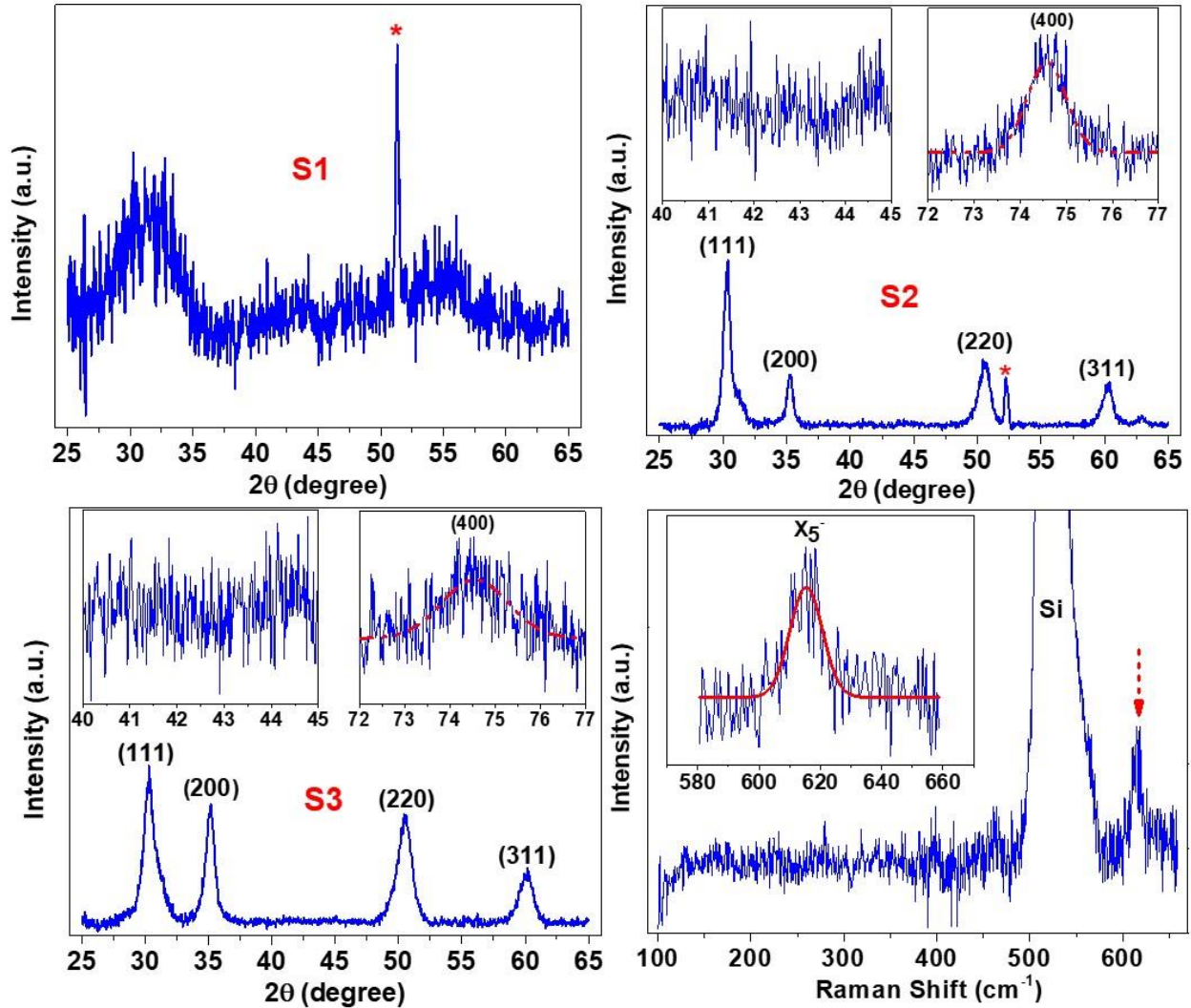

**Figure 2:** XRD patterns of S1, S2 and S3 films; Raman spectra of S3 film. In the XRD patterns of S1 and S2 films, the peak marked by * corresponds to Silicon substrate. The arrow in the Raman spectra shows the cubic phase $X_5^-$ band (which is also shown in the inset).

In agreement with the XRD results, Raman spectroscopy has also indicated the formation of only the cubic phase in the S3 film. The Raman spectra of S3 is shown in Figure 2 – the characteristic cubic phase peak at ~ 615 cm$^{-1}$ ($X_5^-$ mode) [16,39] is clearly visible, whereas the bands corresponding to the tetragonal phase (150 cm$^{-1}$, 260 cm$^{-1}$, 465 cm$^{-1}$, 640 cm$^{-1}$) [39] and monoclinic phase (180 cm$^{-1}$, 190 cm$^{-1}$, 335 cm$^{-1}$, 380 cm$^{-1}$, 475 cm$^{-1}$, 635 cm$^{-1}$) [39] are absent.

EXAFS measurements are then employed to probe the local structural environment and to understand the origin of the cubic phase in the films (S3, S2). The Fourier transform of experimental EXAFS functions ($k^2\chi(k)$) is used to generate the $\chi(R)$ versus R spectra where $\chi(k)$ has been obtained from energy dependent absorption function $\chi(E)$ using the relation $k = \sqrt{\frac{2m(E-E_0)}{\hbar^2}}$, where $m$ is the mass of electron, $E_0$ is the absorption edge energy and $\hbar$ is the Planck's constant. The $k$-range of $3 - 8.5$ Å$^{-1}$ is used for the Fourier transform. Figure 3 shows the best fit of the magnitude and real component of the Fourier transform of Zr K-edge EXAFS function; the fitting is performed using cubic zirconia structure in the $R$-space range of $1 - 3.6$ Å. The local structural parameters, i.e. coordination numbers, bond distances and Debye-Waller (DW) factor, as determined by these fittings are listed in Table 2. In bulk cubic zirconia, zirconium has 8 nearest neighboring oxygen atoms at a mean distance of 2.204 Å followed by 12 zirconium next nearest neighbor atoms at a mean distance of 3.599 Å [20]. It is immediately apparent that the S3 film, which exhibits long range cubic order, shows appreciable deviation from bulk cubic zirconia local structure – the nearest oxygen shell has a much lower coordination number (of around 5.9) and longer bond length (of 2.233 Å) in comparison to bulk cubic zirconia. The decreased number of oxygen neighboring atoms surrounding zirconium indicates the presence of a large number of oxygen vacancies in the system. It is important to understand here that the decreased first shell coordination number does not invalidate the measured stoichiometry of the S3 film. This is because some oxygen atoms may be located at nano-crystalline boundaries, interstitials, film surface etc., instead of their usual lattice sites, and are hence not observed by the short range order EXAFS technique. On the other hand, even these disorderly located oxygen atoms could be detected by RRBS leading to the measured stoichiometry. The disorderly located oxygen atoms leave behind vacancies in their respective lattice positions which are detected by EXAFS.

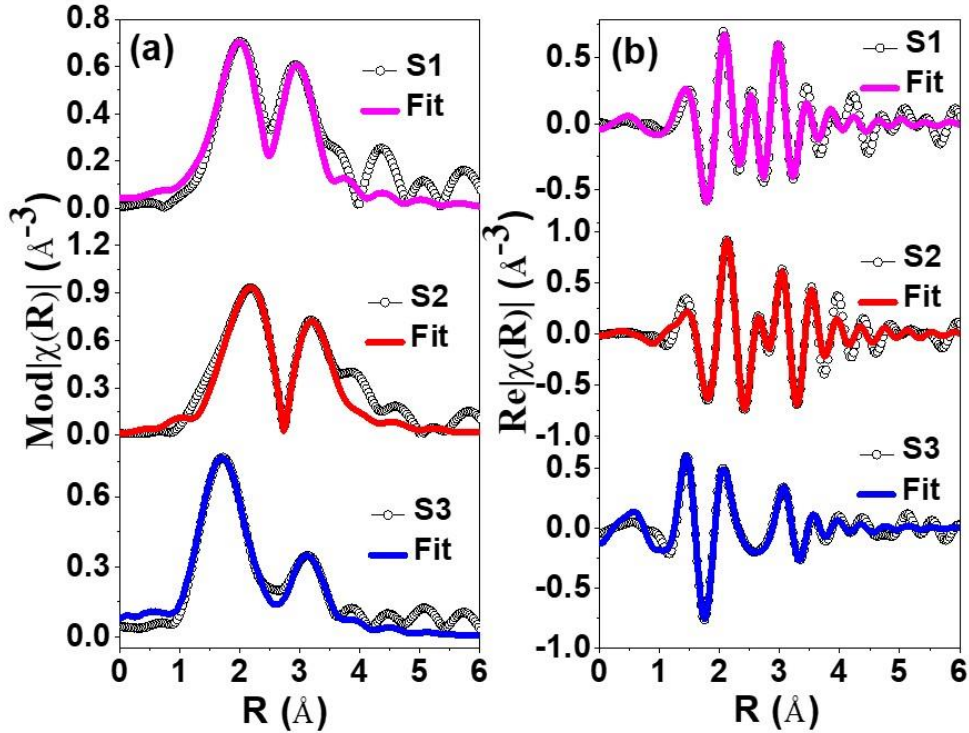

**Figure 3:** (a) Magnitude and (b) real component of the Fourier transform of EXAFS functions ($k^2\chi(k)$) for the films. Dotted lines represent the experimental data whereas the solid lines represent the best fit.

| Sample | $CN_{Zr-O}$ | $R_{Zr-O}$ (Å) | $\sigma^2_{Zr-O}$ (Å$^2$) | $CN_{Zr-Zr}$ | $R_{Zr-Zr}$ (Å) | $\sigma^2_{Zr-Zr}$ (Å$^2$) |
|---|---|---|---|---|---|---|
| S1 | 3.9 (5) | 2.315 (3) | 0.0112 (4) | 8.9 (5) | 3.414 (3) | 0.0069 (3) |
| S2 | 5.4 (4) | 2.413 (4) | 0.0062 (3) | 10.3 (4) | 3.584 (4) | 0.0045 (5) |
| S3 | 5.9 (4) | 2.233 (3) | 0.0098 (3) | 7.5 (4) | 3.542 (3) | 0.0099 (4) |

**Table 2:** Structural parameters obtained from fitting of the experimental EXAFS data at the Zr K-edge. The coordination numbers is CN, R is the bond distance and $\sigma^2$ is the Debye-Waller (DW) factor. The numbers in parentheses indicate the uncertainty in the last digit.

Analogous to the S3 film, the S1 and S2 films also show notable deviation (lower first shell coordination number and longer bond length) from bulk cubic local structure. Interestingly, the first shell (Zr-O) coordination number, and hence the oxygen vacancies, is similar for S2 and S3; while the first shell coordination number is significantly lower for the S1 film i.e. more oxygen vacancies are present in S1. These results positively correlate the amorphous nature of S1 film as opposed to the formation of the cubic phase in S3 and S2 films thus suggesting the role/influence of oxygen vacancies in the cubic phase stabilization mechanism. Note that the DW factor (for the first coordination shell) is also higher for S1, as compared to S3 and S2, thus indicating it as of amorphous nature.

To assess the impact of the presence of the oxygen vacancies on the stability of the cubic phase at the atomistic level, first-principles modeling under the framework of DFT is carried out (see section 2.3 for methodology). We have first given special attention to validate whether the chosen nanoclusters (as shown in Figure 4) are sufficient in reproducing the films under investigation or not. The size of the nanoclusters is taken in such a way that it can reproduce the thinner and thicker films respectively. Here we have considered two nanoclusters of 46 and 131 atoms ($Zr_{14}O_{32}$ and $Zr_{43}O_{88}$) respectively (see Figure 4a and 4b) to represent pristine (without vacancies) thin and thick films respectively. Following this, oxygen vacancies are created into the clusters (see the Figure 4c and 4d corresponding to the S2 and S3 films with oxygen vacancies). We have checked that the clusters first nearest neighbor distance (Zr-O) (as calculated from pair correlation function (PCF) of the nano-clusters) are matching well with the corresponding Zr-O bond length as measured from EXAFS for S2/S3 films. The PCF (EXAFS) values for Zr-O bond distance are 2.41 (2.413) Å and 2.20 (2.233) Å for $Zr_{14}O_{31}$ (S2) and $Zr_{43}O_{86}$ (S3) respectively. The similarity in the structural parameters of the nanoclusters and films are validation of the fact that the nanostructures considered here are sufficiently good to represent the S2/S3 films. Next, to determine the most stable phases, we have drawn the thermodynamic phase diagram by minimizing Gibbs free energy of formation ($\Delta G_f$) of all the configurations using the *ab initio* atomistic thermodynamics approach [34,40]. Note that since the electronic charge can significantly stabilize the oxygen vacancy site, we have considered oxygen vacancy with different charge states (0, -1, -2). The formation energy (Gibbs free energy) of a particular configuration is calculated with respect to different charges (*q*) using the formula

$$\Delta G_f \square^q = E_{tot} (Zr_xO_{y-n})^q - E_{tot} (Zr_xO_y)^0 + n\mu_o + q\mu_e \quad \ldots\ldots (1)$$

where, $E_{tot}(Zr_xO_{y-n})^q$ and $E_{tot}(Zr_xO_y)^0$ are the total energies of the nanocluster with oxygen vacancies in charge state *q* and neutral pristine nanostructure respectively; *n* is the number of oxygen vacancies in the clusters; $\mu_o$ (= $\Delta\mu_o + \frac{1}{2}[E_{O_2} + \frac{h\nu_{OO}}{2}]$) is the chemical potential of oxygen molecule. We determine $\Delta\mu_o$ as a function of *T* and $p_{O_2}$ using equation (2) (see details of this methodology in Ref. [34,40])

$$\Delta\mu_O(T, p_O) = \frac{1}{2}\left[-k_BT \ln\left[\left(\frac{2\pi m}{h^2}\right)^{\frac{3}{2}}(k_BT)^{\frac{5}{2}}\right] + k_BT \ln p_{O_2} - k_BT \ln\left(\frac{8\pi^2 I_A k_BT}{h^2}\right) + \right.$$
$$\left. k_BT \ln\left[1 - \exp\left(-\frac{h\nu_{OO}}{k_BT}\right)\right] - k_BT \ln \mathcal{M} + k_BT \ln \sigma\right] \quad \ldots\ldots (2)$$

where m is the mass, $\mathcal{M}$ is the spin multiplicity, $\nu_{OO}$ is the O–O stretching frequency, and $\sigma$ is the symmetry number. $\mu_e$ is the chemical potential of the electron referenced to the highest occupied molecular orbital (HOMO) of pristine nanostructure. The maximum limit of $\mu_e$ is considered to be the band gap value (~ 5.8 eV) [41-43] of zirconia. Equation (1) allows to directly plot the Gibbs free energy of formation for each configuration as a function of the chemical potential of the electron ($\mu_e$) at finite temperature and pressure (T=300K, $p_{O_2}$ = 1atm, $\Delta\mu_o$ = -0.27 eV), as shown in Figure 5. This yields a plane in the phase diagram corresponding to each charged state, and at any $\mu_e$, the configuration with the lowest lying plane in the phase diagram is the most stable one. From Figure 5 we can see that, irrespective of the size of the cluster, the configuration with neutral O-vacancies (red colored line) is more stable as compared to its pristine structure (dotted black line).

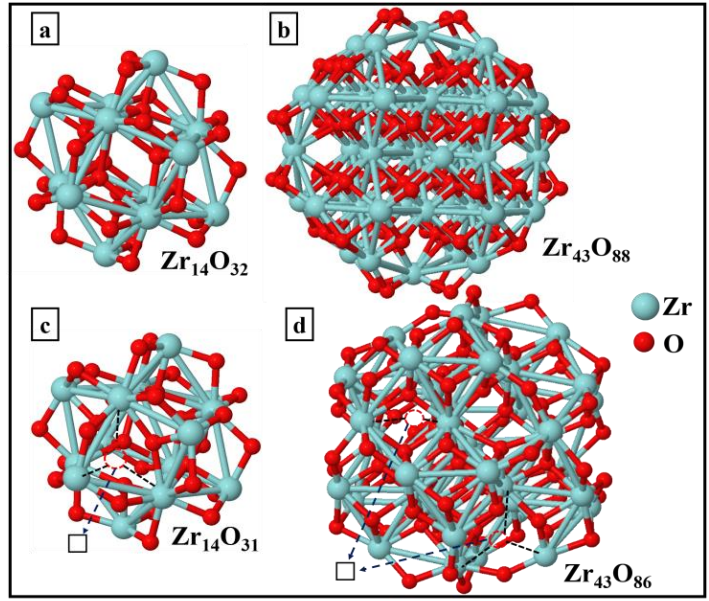

**Figure 4:** Ball and stick model of pristine nanoclusters with different sizes (a) $Zr_{14}O_{32}$, (b) $Zr_{43}O_{88}$ and O-deficient nanoclusters (c) $Zr_{14}O_{31}$, (d) $Zr_{43}O_{86}$. The symbol □ corresponds to the oxygen vacancy in the nanoclusters.

Furthermore, by including the charged configurations in the phase diagram, we have noted that in both the cases the negatively charged state ($q = -2$) exhibits the lowest Gibbs free energy of formation near the lowest unoccupied molecular orbital (LUMO) level around the ambient condition of oxygen gas.

In Figure 5a, with p-type doped condition (i.e. $\mu_e$ close to the HOMO level) oxygen vacancies with −1 charge is the most stable. For other values of $\mu_e$ (i.e. intrinsic/n-type doped) oxygen vacancies with −2 charge is the most favorable state in the phase diagram. From Figure 5b, we see that neutral oxygen vacancies is the most stable configuration for p-type doping whereas at higher values of $\mu_e$ (i.e. close the LUMO level (n-type doping)) oxygen vacancies with −2 charge is the most preferable phase in the phase diagram. Note that oxygen requires two electrons to complete its valence shell. When oxygen vacancy is created, it creates deep donor level where $\mu_e$ usually takes a value near LUMO level (i.e. n-type doping condition). Therefore, in our system, oxygen vacancy is expected to trap electrons to get stabilized. Neutral oxygen vacancy may not be thus found in the system.

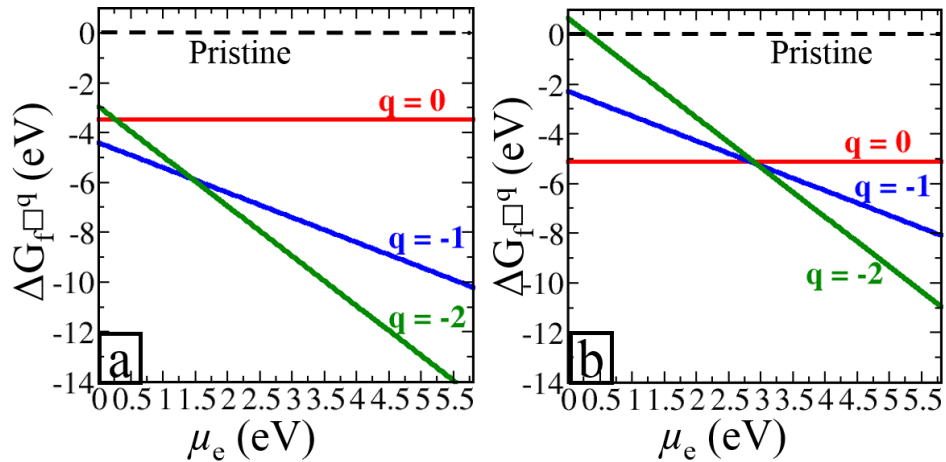

**Figure 5:** The formation energy of different configurations for oxygen deficient nanoclusters ((a) $Zr_{14}O_{31}$, (b) $Zr_{43}O_{86}$) as a function of electron chemical potential $\mu_e$ for different charge states $q$.

This prediction of the oxygen vacancies to be the charged ones has been validated experimentally by photoluminescence (PL) measurements (not shown here). The PL measurement indicates the presence of F and/or $F^+$ color centers i.e. oxygen vacancies with charge states of -2 and/or -1 respectively, in the S3 film.

We, therefore, observe that cubic zirconia with oxygen vacancies is more stable as compared to its pristine structure under ambient conditions. In view of this, we can state that oxygen vacancy in the films is promoting the enhanced stability of the cubic phase.

## 4. Discussion

The formation of cubic phase (under ambient conditions) in the zirconia films without the use of any chemical stabilizers and/or dopants can be attributed to a combination of two mechanisms: first, the 'nanosize' effect and secondly, the presence of oxygen vacancies. The control of the crystallite/grain size has proved to be an efficient way to control the phase stability of nano-materials under ambient conditions. According to the Young- Laplace equation [15,16], the excess pressure *p* inside a crystallite is given by *p = 2γ/r*, where *γ* and *r* are the surface energy and radius of the crystallite respectively. A decrease in the crystallite size thus leads to an increase in their internal pressure and hence favors the stability of phases which would only be stable at high pressure. In general, cubic zirconia is stable at high pressure. Therefore, for the S3 film, in the initial nucleation stages, the zirconia crystallizes in the cubic phase due to the highly compressive internal pressure inside the small crystallites. However, as the cubic crystallites grew larger (as the deposition continued) there would be a decrease in the internal grain/crystallite pressure. Under ambient conditions and in the absence of any chemical stabilizers or dopants, this should have resulted in the transformation of the cubic phase to the more thermodynamically favorable tetragonal or monoclinic phase via the tetragonal distortion of the oxygen sub-lattice (soft $X_2^-$ mode of vibration). In-fact, the critical size for the phase transition from cubic to tetragonal phase, in stabilizer free zirconia, has been found to be ~ 2 nm [22]. Moreover, it has also been reported that the critical size for 100% tetragonal phase stabilization in nano-crystalline zirconia is ~ 6 nm above which zirconia nano-crystallites exist in a core-shell structure with the tetragonal phase as the core and the monoclinic phase as the shell [44]. The crystallite size in our $ZrO_2$ films (S3) is greater than both of these reported critical sizes; however, as discussed in the previous section, formation of only the cubic phase was observed in the films. Thus, the 'nanosize' effect alone cannot explain the formation and stability of the cubic phase, at ambient conditions, for the S3 films.

The presence of oxygen vacancies may be responsible for the stabilization of cubic zirconia with larger crystallite sizes. It is well known that the soft $X_2^-$ mode of vibration of the oxygen atoms mediates the cubic – tetragonal phase transition. The absence of oxygen atoms from its lattice positions (i.e. the presence of oxygen vacancies) results in the breaking of O – Zr bonds. This change in the local structure, i.e. local distortions, hinders the movement of the oxygen atoms especially in the soft $X_2^-$ vibration mode (ultimately suppressing it) and hence locks the crystal in the cubic phase. The stabilization mechanism can be understood from the perspective of phonon dispersion [16]. A vital difference in the phonon dispersion of the cubic and tetragonal phases is the presence of imaginary frequencies around the X high symmetry point [16,45,46] of the cubic phase which indicates its instability. The minimum at the X point is called the $X_2^-$ mode (that connects the cubic to tetragonal phase transition) [16,47]. It is therefore essential to eliminate the imaginary vibration frequencies caused by the oxygen atoms to stabilize the cubic phase (i.e. prevent it from going to the tetragonal phase). Mediated by the oxygen vacancies, the movement of the oxygen atoms is hindered and the soft $X_2^-$ vibration mode is suppressed/destroyed. Thus, the initial cubic $ZrO_2$ nano-crystallites can remain in the cubic phase while growing larger (than the critical 2 nm size) under ambient conditions. Therefore, oxygen vacancies (and the resulting structural distortions) play the key role in stabilizing the cubic phase, without any chemical stabilizers or dopants, under ambient conditions.

The role of oxygen vacancies in stabilizing the cubic phase is also supported, from the point of view of energetics of formation, by our first-principles modelling results which show that, under ambient conditions, cubic zirconia with oxygen vacancies is more stable as compared to its pristine counterpart.

The amorphous nature of the S1 film in-spite of the presence of oxygen vacancies is then quite surprising. Based on the fact that the oxygen co-ordination number for the S1 film is lower as compared to the S3 film, one possible explanation for the amorphous nature of S1 might be that the number of oxygen atoms (in the unit cell) has not reached the threshold value required for crystalline phase formation. In other words, the amount of oxygen vacancies is so high that an overall stable crystalline ordering cannot be formed and the material remains in amorphous (highly disordered) state. The DW factor, which indicates local disorder, is indeed higher for S1 compared to S3 for the Zr-O bond. Thus, for the S1 film, it can be assumed that in the initial stages of the nucleation process, the cubic phase was formed because of the 'nano-size' effect. However, as the final deposition time is much smaller (lower thickness), the amount of oxygen vacancies remains higher in comparison to S3 and the film goes into the amorphous phase. In other words, the S1 film did not get enough time to reach a value of oxygen vacancies required to stabilize the cubic phase. Upon annealing of this film, the required atomic sites are filled with oxygen which eventually leads to a cubic phase (S2 film). As mentioned before, the first shell (Zr-O) coordination number, and hence the oxygen vacancies, is similar for S2 and S3 films. It is to be noted here that although the oxygen (first shell) co-ordination number increases upon annealing, the overall stoichiometry (as determined by RRBS) does not change. This suggests that oxygen atoms have relocated to lattice sites from grain (crystalline) boundaries, inter-granular regions, surface etc. and external oxygen has not been incorporated into the film during annealing.

A critical amount of oxygen vacancies may thus be required in the formation and stabilization of the cubic phase (under ambient conditions) and would be an interesting point to investigate in future studies.

## 5. Conclusions

In conclusion, we have synthesized zirconia films using electron beam evaporation technique that are stable in the cubic phase under ambient conditions without the presence of any chemical stabilizers and/or 'dopants'. Cubic phase was obtained in the as-deposited thicker films (660 nm) and annealed thinner films (140 nm). EXAFS measurements reveal that both of the films have similar amount of oxygen vacancies. Our theoretical analysis show that the cubic structures with oxygen vacancies are more stable than its pristine counterpart (under ambient conditions). These oxygen vacancies are also found to be charged vacancies. The cubic phase stabilization is attributed to these oxygen vacancies that results in the breaking of the O-Zr bond. The latter hinders/suppresses the movement of the oxygen atoms in the soft $X_2^-$ mode of vibration. Due to the suppression of this mode, the cubic phase gets stabilized. Finally, these nano-crystalline cubic zirconia films possess favorable physical properties compared to chemically stabilized cubic zirconia and thus have potential for various technological applications.

## Acknowledgements

We express gratitude towards Dr. D. Kabiraj and Mr. S.R. Abhilash of Target Lab, IUAC New Delhi for their help during the synthesis of the films. We sincerely acknowledge Dr. Sunil Kumar of the Dept. of Physics, IIT Delhi for photoluminescence measurements. P. K. and S. S. are thankful to MHRD, India and CSIR, India respectively for financial assistantship [grant no. 09/086(1231)2015-EMR-I (for S.S.)]. S. B. acknowledges the financial support from YSS-SERB research grant, DST, India (grant no. YSS/2015/001209). S. G. acknowledges the financial support from BRNS, India (grant no. 37(3)/14/25/2017-BRNS/37222). XRD (Dept. of Physics) and Raman spectroscopy (NRF) facilities of IIT Delhi are acknowledged. We also acknowledge the High Performance Computing facility at IIT Delhi for computational resources.